\documentstyle{mn}

\include{epsf}

\begin{document}

\def\lta{\mathrel{\spose{\lower 3pt\hbox{$\mathchar"218$}}
        \raise 2.0pt\hbox{$\mathchar"13C$}}}
\def\gta{\mathrel{\spose{\lower 3pt\hbox{$\mathchar"218$}}
        \raise 2.0pt\hbox{$\mathchar"13E$}}}

\title{Structure and properties of transition fronts in accretion discs}

\author[K. Menou, J.-M. Hameury and R. Stehle]{Kristen Menou$^{1,2}$, 
Jean-Marie Hameury$^3$ and Rudolf Stehle$^4$
\\$^1$ Harvard-Smithsonian Center for Astrophysics, 60 Garden Street,
        Cambridge, MA 02138, USA, kmenou@cfa.harvard.edu
\\$^2$ UPR 176 du CNRS, D\'epartement d'Astrophysique Relativiste et de
        Cosmologie, Observatoire de Paris, Section de Meudon,
\\      F-92195 Meudon C\'edex, France
\\$^3$ UMR 7550 du CNRS, Observatoire de Strasbourg, 11 rue de l'Universit\'e,
        F-67000 Strasbourg, France, hameury@astro.u-strasbg.fr
\\$^4$ Astronomy Group, University of Leicester, Leicester LE1 7RH, U.K., 
        rst@star.le.ac.uk}

\maketitle

\begin{abstract}
We use high-resolution time-dependent numerical simulations of
accretion discs around white dwarfs to study the structure and
properties of transition fronts in the context of the thermal-viscous
disc instability model.  The thermal structure of cooling and heating
fronts is dominated by radiative cooling and viscous heating,
respectively, except in a very narrow precursor region in heating
fronts where advection and radial transport of energy
dominate. Cooling fronts are much broader than heating fronts, but the
widths of both types of fronts scale with the local vertical scale
height of the disc. We confirm that during a fair fraction of the
propagation time of a cooling front, the structure of the inner disc
is close to self-similar. The speed of heating fronts is $\sim$ a few
km s$^{-1}$, while the speed of cooling fronts is $\sim$ a fraction of
a km s$^{-1}$. We show that direct measurements of the speed of
transition fronts probably cannot discriminate between various
prescriptions proposed for the viscosity parameter $\alpha$.  A
natural prediction of the disc instability model is that fronts
decelerate as they propagate in the disc, independent of the
prescription for $\alpha$. Observation of this effect would confirm
that dwarf nova outbursts are driven by the thermal-viscous
instability. Most of our results also apply to low mass X-ray binaries
in which the accreting object is a neutron star or a black hole.
\end{abstract}

\begin{keywords}
accretion, accretion discs -- instabilities -- novae, cataclysmic variables
-- binaries : close
\end{keywords}

\section{Introduction}

It is widely accepted that the disk instability model (DIM) provides
the correct description of dwarf nova outbursts (e.g Bath \& Pringle
1982, Smak 1984, Lin, Papaloizou \& Faulkner 1985, Cannizzo 1993a) and
probably of soft X-ray transient events (van Paradijs \& Verbunt 1984,
Mineshige \& Wheeler 1989, Cannizzo 1998a).  In this model, the
variability of a steadily fed thin accretion disc is due to the
occurrence of a thermal-viscous instability in the disc. For a certain
range of mass transfer rates from the secondary, $\dot{M}_T$, the
effective temperature of a disc annulus lies within the unstable range
$5000-8000$ K, which corresponds to hydrogen recombination inside the
disc. The unstable annulus then experiences a limit cycle during which
matter is processed at rates larger or smaller than the mean rate
$\dot{M}_T$ (see Cannizzo 1993b or Osaki 1996 for reviews).

Transition fronts are essential to the DIM because they are the link
between the local instability of an annulus and the global evolution
of the disc. The time required by a heating front to cross the disc is
related (depending on wavelength) to the rise time
of an outburst. Similarly, the time required for a cooling front to
cross the disc defines the decay time of an outburst in the DIM (if
the entire disc is brought to the hot state, the decay time is
increased by a ``viscous plateau'' phase, see Cannizzo 1993a).
Observations of the rise, decay and recurrence times therefore allow a
direct test of the predictions of the DIM.

Several attempts to determine the structure and properties of
transition fronts exist in the literature. Analytical studies lead
only to qualitative results or to complex results that require
numerical simulations for calibration (Meyer 1984, Lin et al.  1985,
Vishniac \& Wheeler 1996, Vishniac 1997). On the other hand, numerical
studies have difficulties in correctly handling transition fronts
because the fronts are very narrow and their propagation involves time
and length scales which vary by several orders of magnitude in a
single outburst. As discussed by Hameury et al. (1998, hereafter
HMDLH), heating fronts are unresolved in most numerical simulations
while cooling fronts are often barely resolved, in particular close to
the disc inner edge (see e.g. Cannizzo, Chen \& Livio 1995). The poor
resolution of transition fronts not only affects localized regions of
the disc but it has also consequences for the global evolution of the
disc (Cannizzo 1993a, HMDLH).

HMDLH have recently developed an implicit time-dependent numerical
code which solves the disc equations on an adaptive grid. The code
allows the resolution of very narrow structures in the disc at a
relatively low computational cost. Here we use this code to study the
detailed structure and the propagation velocities of heating and cooling
fronts. We show that the width of any transition front is proportional
to the disc pressure scale height $H$, and not to $(HR)^{1/2}$ as
proposed by Cannizzo et al. \shortcite{ccl95}.  We confirm the
existence of a self-similar regime of the hot inner disc during the
propagation of a cooling front, and we discuss the possible
reflections of cooling (resp. heating) fronts into heating
(resp. cooling) fronts.

\section{Disc instability model and transition fronts}

\subsection{Disc equations}

The time evolution of an accretion disc is driven by thermal-viscous
processes. {For a Keplerian disc t}he viscous equation, describing mass and 
angular momentum conservation, can be written as
\cite{l74,p81}:
\begin{equation}
\frac{\partial \Sigma}{\partial t}=\frac{3}{R}
\frac{\partial}{\partial R} \left[ R^{1/2} \frac{\partial}{\partial R}
\left( \nu \Sigma R^{1/2} \right) \right],
\end{equation}
where $\nu$ is the kinematic viscosity,
$\Sigma$ is the surface density of the disc, and $R$ is the
distance from the central object. The viscosity is parameterized
according to $\nu = \alpha c_{\rm s} H$, where $c_{\rm s}$ is the
sound speed in the disc midplane, $H$ is the vertical scale height of the
disc and the parameter $\alpha$ ($\leq 1$) describes our ignorance of
the viscous processes. Shakura and Sunyaev (1973) first proposed this
prescription with $\alpha$ constant, but it was later shown by Smak
(1984) that $\alpha$ must be larger for the hot ionized disc than for
the cold neutral disc in order to obtain large amplitude outbursts in
the DIM. 

In the following, we use two parameterizations often adopted in the
literature. First, we adopt an $\alpha_{\rm hot}-\alpha_{\rm cold}$
prescription and an arbitrary functional dependence of $\alpha$ with
the disc central temperature $T_c$ which is similar to that used in
HMDLH, i.e.
\begin{eqnarray}
\log (\alpha)=\log(\alpha_{\rm cold})& +& \left[ \log(\alpha_{\rm hot})-
\log( \alpha_{\rm cold} ) \right] \nonumber \\ & &
\times \left[1+ \left( \frac{2.5 \times 10^4 \; \rm K}{T_{\rm c}} \right)^8
\right]^{-1},
\label{eq:alpha}
\end{eqnarray}
where $\alpha_{\rm hot}$ and $\alpha_{\rm cold}$ are the two constant
values of $\alpha$ on the hot and the cold stable branches of the thermal
equilibrium curves (or S-curves), respectively.

The second parameterization we consider is 
\begin{equation}
\alpha = \alpha_0 (H/R)^{1.5},
\end{equation}
where $\alpha_0$ is a constant ($\alpha_0=50$ here). Such a dependence
has been strongly advocated by several authors to explain the observed
exponential decays of soft X-ray transients \cite{ccl95,vw96}. This
interpretation may face some difficulties, however (\S 4.6). Note
that the viscosity parameter $\alpha \propto R^{3/4}$ in this case
(e.g.  Mineshige \& Wheeler 1989).

The importance of allowing the disc outer edge, $R_{\rm out}$, to vary
with time in order to reproduce correct outburst cycles was emphasized
by HMDLH and before by Smak (1984) and Ichikawa and Osaki (1992).
Transition fronts are, however, localized structures and should not be
influenced by the physical conditions at the disc outer edge. This is
confirmed by a comparison between models differing only in whether or
not $R_{\rm out}$ varies with time, which shows that the structure and
the properties of the fronts are identical in the two classes of models
(except when a front reaches $R_{\rm out}$). In the following, we
assume that $R_{\rm out}$ is fixed.

The thermal equation (describing energy conservation; Faulkner, Lin \&
Papaloizou 1983, HMDLH) is given by:
\begin{equation}
{\partial T_{\rm c} \over \partial t} = { 2 (Q^ + -Q^- + J) \over C_P
\Sigma} - {\Re T_{\rm c} \over \mu C_P} {1 \over R} {\partial (R
V_{\rm r}) \over \partial R} - V_{\rm r} {\partial T_{\rm c} \over
\partial R},
\label{eq:heat}
\end{equation}
where $Q^+$ and $Q^-$ are the vertically-averaged rates of viscous
heating and radiative cooling, respectively, $\Re$ is the perfect gas constant,
$C_P$ is the specific heat at constant pressure of the gas, $\mu$ its
mean molecular weight and $V_{\rm r}$ its radial velocity (positive
outward). The second term on the r.h.s. of Eq.~(\ref{eq:heat})
represents the contribution from pressure work and will be referred
to as $Q_{\rm pdv}$. The third term ($Q_{\rm adv}$)
represents the contribution from energy advection. The term $J$ (or
equivalently $Q_j$) accounts for the radial transport of energy
(viscous or radiative). The (viscous) flux carried by turbulent eddies
with a characteristic velocity $v_{\rm e}$ and size $l_{\rm e}$ is
obtained dimensionally in the framework of the $\alpha$-prescription:
\begin{equation}
F_{\rm e} = C_P \Sigma v_{\rm e} {\partial T_{\rm c} \over \partial R}
l_{\rm e} = {3 \over 2} \nu C_P \Sigma {\partial T_{\rm c} \over
\partial R},
\label{eq:fturb}
\end{equation}
and $J$ is then given by:
\begin{equation}
J = \frac{1}{R} \frac{\partial}{\partial R} \left( R F_{\rm e}
\right).
\label{eq:j}
\end{equation}
A similar expression is obtained for the radiative transport of energy
\cite{lpf85}. Other prescriptions were proposed for $J$ but they give
results which are similar, in terms of outburst cycles, to those
obtained with Eqs.~(\ref{eq:fturb}) and~(\ref{eq:j}) (HMDLH; see also
\S 6).

Simulations show that most of the disc remains close to thermal
equilibrium during its time evolution. In that case, the thermal
equation basically reduces to $Q^+ = Q^-$. Transition fronts
correspond, however, to regions far from thermal equilibrium in which
the other terms in Eq.~(\ref{eq:heat}) are no longer negligible.

\subsection{Numerical models}

We use the numerical code developed by HMDLH with 800 grid
points. This corresponds to about 100 points in the transition fronts,
and is more than sufficient to avoid resolution-limited results.

All the models predict strictly periodic sequences of outbursts with
(eventually) various amplitudes (see e.g. Fig.~8 of HMDLH). The
sequences differ from one model to another (in number and amplitude of
the outbursts). In the following, we refer to such a sequence as the
outburst cycle of a model.

The disc outer radius is fixed to $R_{\rm out}=4 \times 10^{10}$ cm.
The disc inner radius, $R_{\rm in}=5 \times 10^{8}$ cm, is equal to
the white dwarf radius of mass $M_1=1.2~M_{\odot}$. A comparison of
models with various mass transfer rates ($\dot{M}_T$) shows that,
although the outburst cycles differ from one model to another, the
intrinsic structure and properties of transition fronts do not depend
on $\dot{M}_T$. A value $\dot{M}_T=10^{-9}~M_{\odot}~{\rm yr}^{-1} =
6.66 \times 10^{16}$ g s$^{-1}$ is used in all models presented here.

We have considered four models to investigate the influence of the
viscosity prescription on the structure and the properties of
transition fronts.  In three of them, we use an $\alpha_{\rm
hot}-\alpha_{\rm cold}$ prescription: $\alpha_{\rm hot}=0.1$ and
$\alpha_{\rm cold}=0.02$ (model {\it h0.1.c0.02}), $\alpha_{\rm
hot}=0.2$ and $\alpha_{\rm cold}=0.02$ (model {\it h0.2.c0.02}) and
finally $\alpha_{\rm hot}=0.1$ and $\alpha_{\rm cold}=0.01$ (model
{\it h0.1.c0.01}). The fourth model uses $\alpha=50(H/R)^{1.5}$ (model
{\it 50hr1.5}).

Global stability considerations (e.g. Cannizzo 1993b) and numerical
simulations show that cooling fronts invariably appear in the outer
disc. On the contrary, outbursts may be triggered anywhere in the
disc, and both inside-out and outside-in heating fronts are observed
in simulations (Smak 1984, HMDLH). All outbursts are of the
inside-out type in the models presented here, except the largest
amplitude outbursts of model {\it h0.1.c0.01} and all the outburst of
model {\it 50hr1.5} which are of the outside-in type.

\subsection{Transition fronts}

We define the location of a front, $R_{\rm front}$, as the radius at
which the central temperature $T_c = 1.8 \times 10^4$~K in the
disc. This value of $T_c$ is a signature of the presence of a
transition front because it lies in the range of central temperatures
for which an annulus is thermally and viscously unstable. The speed of
a front, $V_{\rm front}$, is obtained by numerical differentiation of
$R_{\rm front}$.

In the following, the structure of transition fronts is shown each
time for a specific value of $R_{\rm front}$. Simulations show that
the structures of transition fronts are qualitatively independent of
their location in the disc. (In addition, cooling fronts evolve in a
nearly self-similar way.) The general properties of the fronts do not
depend on the viscosity prescription either.

From now on, the various terms $Q/Q^+$ in Eq.~(\ref{eq:heat}) used to
describe the thermal structure of a transition front are defined as
positive if they correspond to a local heating and negative if they
correspond to a local cooling in the disc (i.e. contrary to the
positive convention for $Q^-$ in Eq.~(\ref{eq:heat})).

\section{Heating fronts}

\subsection{Structure of inside-out heating fronts}

\begin{figure}
\epsfysize=8.5cm
\epsfxsize=8.5cm
\begin{displaymath}
\epsfbox{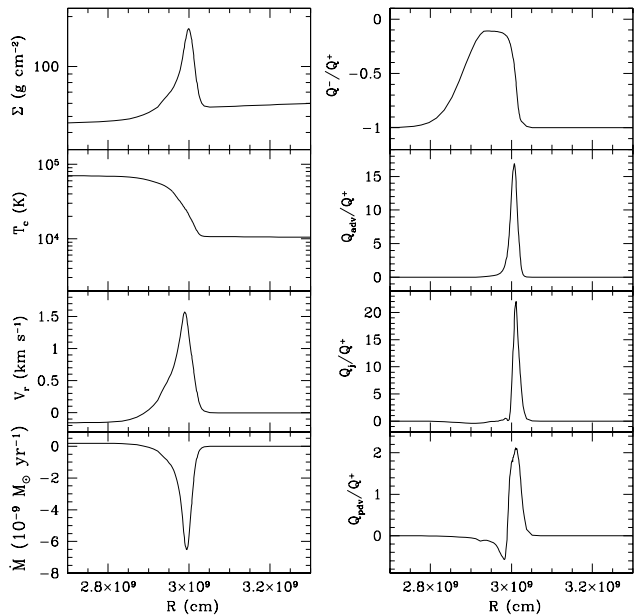}
\end{displaymath}
\caption{Structure of an inside-out heating front located at $R_{\rm
front} \sim 3 \times 10^9~{\rm cm}$ (model {\it h0.1.c0.02}). The
structure is dominated by a strong outflow of gas
and by viscous heating, except in a narrow
precursor region where radial diffusion and advection of energy
dominate.}
\label{fig:inoutheat}
\end{figure}


Figure~\ref{fig:inoutheat} shows the structure of an inside-out
heating front located at $R_{\rm front} \sim 3 \times 10^9~{\rm cm}$
in model {\it h0.1.c0.02}. It is characterized by a strong outflow of
gas which is coincident with a sharp spike in the $\Sigma$ profile and
a steep gradient of $T_c$ (or equivalently viscosity). The outflow and
the formation of the spike in $\Sigma$ were explained by Lin et
al. (1985); they appear because the gas which carries
excess angular momentum from the inner disc reaches colder regions
where transport of angular momentum is reduced.

The dominant term in the thermal equation inside an inside-out heating
front is, by far, the viscous heating term $Q^+$.
Figure~\ref{fig:inoutheat} shows, however, the existence of a
substructure in the front: a precursor region (narrower than the front
itself) in which the energetics of the gas is dominated by advection
and radial transport of energy. The contribution from pressure work is
small but not negligible in the precursor region. The thermal
structure of the front can be understood as follows. The hot gas which
penetrates in colder regions of the disc carries a substantial amount of
heat ($Q_{\rm adv}$). The gas is slowed down where the transport of
angular momentum is reduced and it contracts ($Q_{\rm pdv}$). The
precursor region is the region of strongest gradient of $T_c$ (hence
the large value of $Q_j$).

Note that the region where heating ($Q^+$) dominates over cooling is
broader than the spike in $\Sigma$ and the precursor region
corresponds to the rising part of the spike of $\Sigma$.

\subsection{Structure of outside-in heating fronts}

\begin{figure}
\epsfysize=8.5cm
\epsfxsize=8.5cm
\begin{displaymath}
\epsfbox{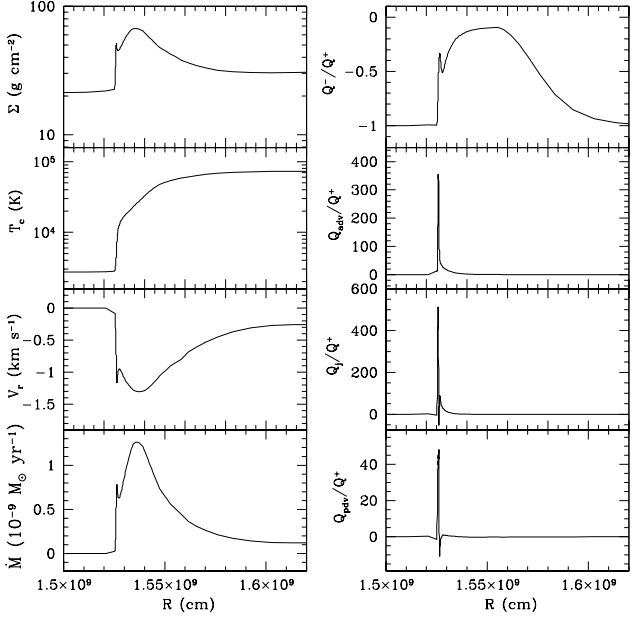}
\end{displaymath}
\caption{Structure of an outside-in heating front located at $R_{\rm
front} \sim 1.53 \times 10^{9}~{\rm cm}$ (model {\it 50hr1.5}). The
structure is dominated by a strong inflow of gas and by viscous
heating, except in a very narrow precursor region where radial
diffusion and advection of heat dominate.}
\label{fig:outinheat}
\end{figure}


Figure~\ref{fig:outinheat} shows the structure of an outside-in
heating front located at $R_{\rm front} \sim 1.5 \times 10^9~{\rm cm}$
in model {\it 50hr1.5}. The structure is expected to be different from
that of an inside-out heating front since the matter crossing the
transition region no longer carries excess angular momentum from the
inner disc. Instead, this matter now corresponds to the bulk of
accretion in the disc, while excess angular momentum is being freely
carried away outward \cite{lpf85}.

The accretion of hot gas is revealed by the strong inflow and the
spike of $\Sigma$ where the gas enters colder regions. Note the very
steep gradient of $T_c$ at the interface between the hot and cold
regions of the disc and the presence of a substructure in the profiles
of $\Sigma$, $V_r$ and $\dot{M}$ (corresponding to the precursor
region).

The thermal structure of an outside-in heating front is similar to
that of an inside-out heating front, despite their opposite direction
of propagation. Viscous heating ($Q^+$) is dominant, except in the
precursor region where advection and radial diffusion of energy
dominate the energetics of the gas. Once again, the pressure work term
($Q_{\rm pdv}$) is small but not negligible in the precursor
region. The precursor region is, however, much narrower (as compared
to the total width of the front) and its structure more complex than
in an inside-out heating front.

\subsection{Heating front velocity}

\begin{figure}
\epsfysize=8.5cm
\epsfxsize=8.5cm
\begin{displaymath}
\epsfbox{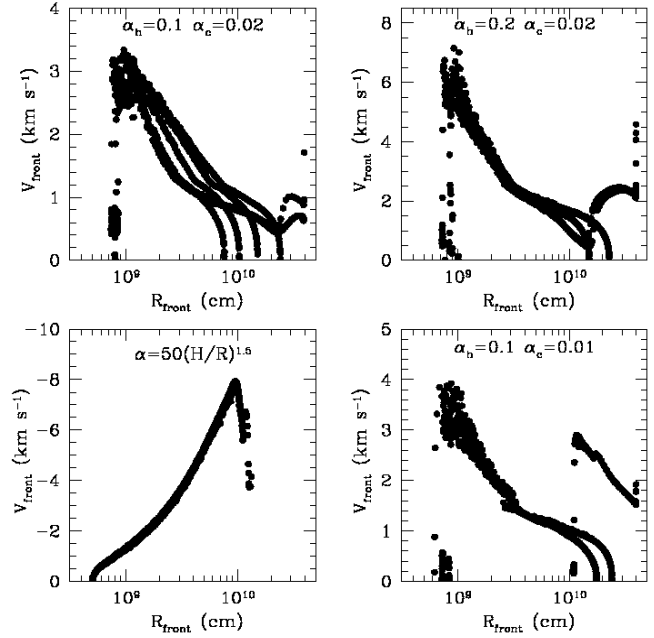}
\end{displaymath}
\caption{The four panels show the speed of successive heating fronts
in our four models with various $\alpha$-prescriptions
(dots). Outside-in (model {\it 50hr1.5}) and inside-out (three other
models) heating fronts slow down during their propagation in the
disc. The speed of heating fronts significantly depends on the value
of $\alpha_{\rm hot}$.}
\label{fig:heatspeedmult}
\end{figure}


Figure~\ref{fig:heatspeedmult} shows the speed of heating fronts for
the four models with various $\alpha$-prescriptions.  Each dot
represents the speed of the front, $V_{\rm front}$, at a specific
location $R_{\rm front}$ in the disc.  Successive outbursts are
triggered at nearly the same ``ignition'' radius during an outburst
cycle in all the models, except in model {\it h0.1.c0.01} for which
the outburst cycle is made of one outside-in (not shown here for
clarity) and several inside-out outbursts. The simulations show that
the front velocities strongly depend on the surface density profile at
the onset of the thermal-viscous instability. Since mass piles up and
diffuses from the outer regions of the disc during an outburst cycle,
the front velocities vary from one outburst to another during the
cycle, as seen from Fig.~\ref{fig:heatspeedmult}.  For comparison, the
sound speed is $\sim 15$ km s$^{-1}$ for the characteristic
temperature $T_c= 18,000$ K in the transition front.

The velocity profiles of outside-in heating fronts (model {\it
50hr1.5}) are very different from those of inside-out heating fronts
(three other models). (This is not just due to the different
$\alpha$-prescriptions.) A common characteristic of the two types of
fronts is, however, a deceleration as the front propagates in
the disc (a characteristic of cooling fronts as well; see \S 4.2).

It is not possible to compare quantitatively the speed of inside-out
and outside-in heating fronts from Fig.~\ref{fig:heatspeedmult}
because a different $\alpha$-prescription is used in the models where
the two types of fronts appear. Fortunately, we obtain both outside-in
and inside-out outbursts in our model {\it h0.1.c0.01}. The maximum
speeds of the two types of fronts in this model are similar ($\sim
3~{\rm km~s^{-1}}$) but outside-in heating fronts cross the disc
faster than inside-out heating fronts because their speed is high over
a larger range of radii. This is in agreement with the fact that
outside-in outbursts are known to produce more asymmetric lightcurves
(i.e. shorter rise times) than inside-out outbursts (e.g. Smak 1984,
HMDLH).

The speed of heating fronts slightly increases when $\alpha_{\rm cold}$
decreases. Increasing the value of $\alpha_{\rm hot}$ results in a
substantial, approximately linear, increase of $V_{\rm front}$ which is
of order of $\alpha_{\rm hot} c_{\rm s}$ as in an ignition front
(Meyer 1984). We find no significant change in the speed of heating
fronts at a given $R_{\rm front}$ when the mass of the central white
dwarf is decreased from $M_1=1.2~M_{\odot}$ to $0.6~M_{\odot}$ in model
{\it h0.1.c0.02}.

It was argued by Lin et al. (1985) that the velocity of a transition
front must be close to the speed of gas in the front. We compared the
speed of heating fronts with the maximum speed of gas inside the
fronts, $V_{\rm r,max}$ (see Figs.~\ref{fig:inoutheat}
and~\ref{fig:outinheat}). The velocity of inside-out and outside-in
heating fronts appears, indeed, close to $V_{\rm r,max}$ in the
fronts. For instance, $V_{\rm front} \sim 3/2~V_{\rm r,max}$ in models
{\it h0.1.c0.02} and {\it 50hr1.5}.  Note that the speed of the cold
gas into which a heating front penetrates is negligibly small compared
to the speed of the front.

\subsection{Width of heating fronts}

The width of a transition front was defined in HMDLH as the region of
the disc over which the variation of the viscosity parameter $\alpha=
\alpha(T_c)$ is 90 \% of the total variation between
$\alpha_{\rm cold}$ and $\alpha_{\rm hot}$. This definition, although
appropriate for an $\alpha_{\rm hot}-\alpha_{\rm cold}$ prescription,
cannot be used when $\alpha$ is a smooth function of $H/R$ (in
particular model {\it 50hr1.5}).

We instead define the transition region as a zone with large thermal
imbalance where the temperature changes on short timescales. More
specifically, if the local timescale of change in temperature is
\begin{equation}
t_{\rm T} (R) \equiv \left| \frac{\partial \ln T_c}{\partial t} (R)
\right| ^{-1},
\end{equation}
we define the transition region as the region of the disc where
$t_{\rm T} (R) < 3 t_{\rm T} (R_{\rm front})$. $R_{\rm front}$ is the
location of the front, as defined previously ($T_c(R_{\rm front})
\equiv 18,000$ K). The factor 3 is arbitrary but it has been chosen to
give results close to what an eye estimate would give. We confirmed
that the above definition leads to widths which are comparable to
those found in HMDLH but also comparable to those found with a
definition of the transition region based on a deviation of the ratio
$| Q^-/Q^+|$ from unity. (This last definition should be equivalent to
the definition used by, e.g., Cannizzo 1994; Cannizzo et al. 1995.)

\begin{figure}
\epsfysize=8.5cm
\epsfxsize=8.5cm
\begin{displaymath}
\epsfbox{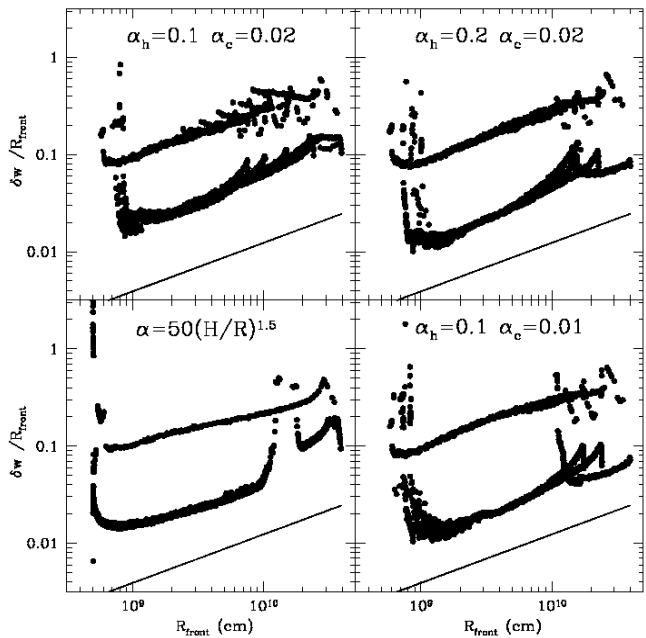}
\end{displaymath}
\caption{The four panels show the fractional width ($\delta w/R_{\rm
front}$) of cooling fronts and heating fronts in our four models with
various $\alpha$-prescriptions (dots). In each panel, the upper line
of dots corresponds to the width of successive cooling fronts and the
lower line of dots corresponds to the width of successive heating
fronts. The solid line shows the fractional scale height of the disc
($H/R_{\rm front}$). Note that heating fronts in model {\it 50hr1.5}
are of the outside-in type.}
\label{fig:dwmult1.2}
\end{figure}


Figure~\ref{fig:dwmult1.2} shows the fractional width ($\delta
w/R_{\rm front}$) of heating and cooling fronts (dots) as a function
of the front location in the four models with various
$\alpha$-prescriptions. In each panel, the upper line of dots
corresponds to cooling fronts and the lower line of dots corresponds
to heating fronts. The width of cooling fronts is further discussed in
\S 4.3.

The width of a heating front is roughly proportional to the local
vertical scale height $H$ (note that $H=c_s/\Omega_K$ is defined at a
fixed value $T_c=1.8 \times 10^4$ K here). This is not unexpected
since, except for the narrow precursor region, the thermal structure
of the front is dominated by the heating term, $Q^+$. Consequently,
assuming that $V_{\rm front}$ varies much less rapidly than $R_{\rm
front}^{3/2}$ (in agreement with our results, \S 3.3), $\delta w$ is
the distance traveled by the heating front in a few thermal
timescales $\tau_{\rm th} \equiv (\alpha \Omega_{\rm K})^{-1}$
(e.g. Frank, King \& Raine 1992), i.e.
\begin{equation}
\delta w \propto V_{\rm front} \tau_{\rm th} \propto (\alpha
c_s/\alpha \Omega_K) \propto H \propto R_{\rm front}^{3/2}.
\label{eq:deltaw}
\end{equation}
These results are consistent with the conclusions of Meyer (1984,
1986) and Papaloizou \& Pringle (1985). 

A comparison of the various panels in Fig.~\ref{fig:dwmult1.2} shows
the influence of $\alpha$ on the width of a heating front. Increasing
$\alpha_{\rm hot}$ or decreasing $\alpha_{\rm cold}$ results in
narrower fronts. It is not possible to interpret this result simply
from Eq.~(\ref{eq:deltaw}), however, because the speed of heating
fronts is also affected in a non-trivial way by changes in the
viscosity parameter $\alpha$ (cf \S 3.3).

In model {\it 50hr1.5}, transition fronts have profiles of $\delta
w/R_{\rm front}$ which are slightly flatter than the solid line
($H/R_{\rm front}$). According to the three other models, reduced
widths are expected for larger values of $\alpha_{\rm hot}$. The
flatter profiles of $\delta w$ are therefore consistent, at least
qualitatively, with the additional variation of $\alpha$ with radius in
this particular model ($\alpha \propto R^{3/4}$).

We also investigated the effect of varying the mass of the central
object on the width of transition fronts. We decreased $M_1$ from
1.2~M$_{\odot}$ to 0.6~M$_{\odot}$ in model {\it h0.1.c0.02}, which
results in a reduced vertical gravity and therefore an increased
vertical scale height $H$ of the disc. At a given $R_{\rm front}$, the
front widths are increased, but the ratio of heating front (or
equivalently cooling front) width to $H$ is reasonably constant. This
is additional evidence for a proportionality between $\delta w$ and
the local thermal timescale in the disc, which depends on $M_1$ via
$\Omega_K$.

\subsection{Pairs of heating fronts}

When the thermal-viscous instability is triggered at a given ignition
radius, two heating fronts appear which propagate in opposite
directions in the disc. Because the ignition radius is often close to
one of the disc edges, one of the two fronts is actually short-lived
(i.e. quickly reaches the disc edge). The other heating front is long
lived, propagates over a large range of radii and determines the
global evolution of the disc.

\begin{figure}
\epsfysize=8.5cm
\epsfxsize=8.5cm
\begin{displaymath}
\epsfbox{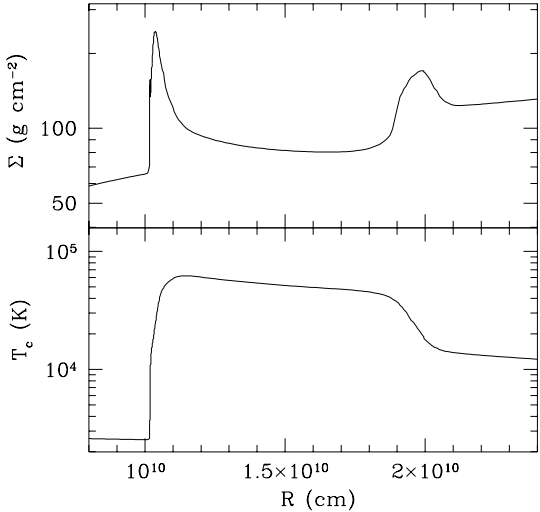}
\end{displaymath}
\caption{Radial profiles of $\Sigma$ and $T_c$ in a situation of
coexistence of an inside-out and an outside-in heating front in the
disc (model {\it 50hr1.5}). Both fronts will significantly contribute
to the global evolution of the disc.}
\label{fig:twinfronts}
\end{figure}


It is possible, however, to obtain models in which a pair of heating
fronts appears somewhere in the middle of the disc.
Figure~\ref{fig:twinfronts} shows such a situation (model {\it
50hr1.5}), where an inside-out and an outside heating fronts appear
around $R \sim 1.5 \times 10^{10}$ cm. In that case, the two fronts
contribute significantly to the global evolution of the disc. We
referred to this type of intermediate outburst as an outside-in
outburst.

More generally, simulations show that the value of the ignition radius
in a specific model depends on the mass transfer accretion rate
$\dot{M}_T$ (HMDLH). This implies that a continuum of lightcurve
shapes is possible, from quasi-symmetric (firm inside-out outburst) to
strongly asymmetric (firm outside-in outburst). Consequently, the
classification in two types of fronts initially proposed by Smak
(1984) must be used with caution.

A careful look at Figure~3 of HMDLH reveals a sudden increase and then
decrease of the inner accretion rate, $\dot{M}_{\rm acc}$, which
precedes the bulk of accretion in inside-out outbursts; this spike is
very short lived, contains almost no energy, and is probably not
observable. This is a signature of the arrival of an in-propagating
heating front at the disc inner edge. This front is the short-lived
companion of the long lived inside-out heating front which propagates
over a large range of radii in the disc. Outside-in outbursts
(e.g. Fig.~4 of HMDLH) do not show this burst of $\dot{M}_{\rm acc}$
because the outside-in heating front which reaches the disc inner edge
is responsible for the bulk of the outburst accretion in that case.

\subsection{Reflection of an inside-out heating front}

The surface density after the passage of a heating front must be
larger than the local value of $\Sigma_{\rm min}$ (the minimum density
on the hot branch of the S-curve) to allow the front to propagate
further in the disc. If not, a cooling front appears where $\Sigma <
\Sigma_{\rm min}$ in the disc (just behind the $\Sigma$ spike). Once
the cooling front develops, the transport of excess angular momentum
from the inner disc is strongly reduced and the propagation of the
heating front is stopped. The spike in $\Sigma$ then diffuses under
the action of viscous processes. This process is usually referred to
as a reflection, although it corresponds more exactly to the
appearance of a new cooling.

In the case of outside-in fronts, matter flows from the outer parts of
the disc, which contain most of the disc mass, so that $\Sigma$
steadily increases at any given radius, and the heating front is
always able to reach the disc inner edge without being reflected in
our models.

The situation is quite different for inside-out heating fronts, for
which mass is transferred from the inner parts to the outer parts of
the disc. The comparatively small amount of mass in the inner disc
leads to several reflections of inside-out heating fronts during one
outburst cycle. After these multiple reflections, an inside-out
heating front can reach $R_{\rm out}$ without being reflected because
enough mass has accumulated in the outer disc and $\Sigma$ behind the
front remains above $\Sigma_{\rm min}$ during the entire propagation.

The reflections also occur in mini-outbursts (as compared to
``normal'' outbursts in which most of the disc -- if not all --
undergo a transition to the hot state) which are found in many
numerical models (see e.g. Cannizzo 1993a, HMDLH). The significance of
these mini-outbursts is unclear because they have probably not been
observed.

\section{Cooling fronts}

\subsection{Structure of cooling fronts}

\begin{figure}
\epsfysize=8.5cm
\epsfxsize=8.5cm
\begin{displaymath}
\epsfbox{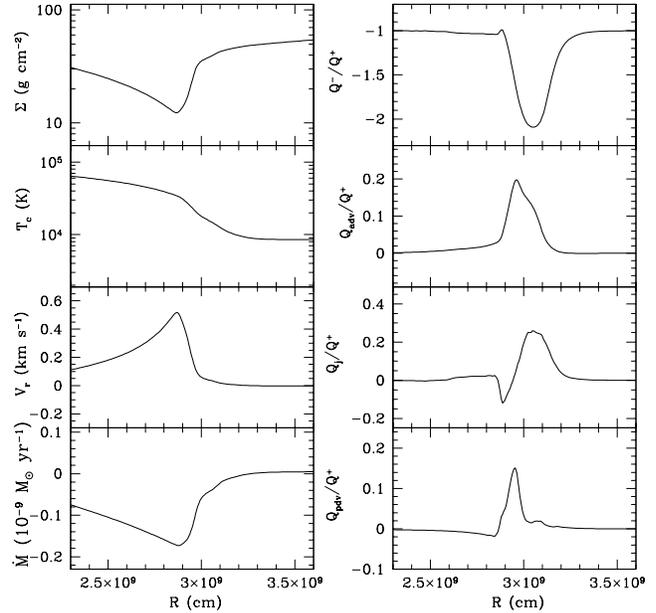}
\end{displaymath}
\caption{Structure of a cooling front located at $R_{\rm front} \sim 3
\times 10^9~{\rm cm}$ (model {\it h0.1.c0.02}). The structure is
dominated by an outflow of gas and by radiative cooling.}
\label{fig:cool}
\end{figure}


Figure~\ref{fig:cool} shows the structure of a cooling front located
at $R_{\rm front} \sim 3 \times 10^9~{\rm cm}$ in model {\it
h0.1.c0.02}. The innermost, hot, regions of the disc accrete at a much
higher rate than the outer cold regions (see e.g. Fig.~9 of Cannizzo
et al. 1995). A substantial amount of angular momentum flows outward
in the inner regions of the disk.  The gas which carries this excess
angular momentum is stopped when it reaches colder regions of the
disc, where the transport of angular momentum is reduced. This results
in a large outflow and an abrupt increase of $\Sigma$ where rapid
cooling sets in, as seen from Fig.~\ref{fig:cool}.

As noted by Vishniac \& Wheeler (1996) (see also Vishniac 1997), the
inner disc is in quasi-steady state during the propagation of a
cooling front; the accretion rate is almost independent of radius and
it slowly decreases with time. The profiles of $\Sigma$ and $T_c$ in
these regions (not shown in Fig.~\ref{fig:cool}, but see, e.g.,
Fig.~5a of HMDLH) are close to those of the steady solution of Shakura
\& Sunyaev (1973). Between the inner quasi-steady disc and the front
itself, there is a relatively broad region (called precursor region by
Vishniac \& Wheeler 1996) where the profiles of $\Sigma$ and $T_c$
deviate substantially from the Shakura-Sunyaev profiles.

Figure~\ref{fig:cool} shows that the thermal structure of a cooling
front is dominated, although not strongly, by radiative cooling
($Q^-$): in first approximation, gas freely cools inside a cooling
front (Vishniac \& Wheeler 1996). The three non-local terms in the
energy equation, namely $Q_{\rm adv}$, $Q_{\rm j}$ and $Q_{\rm pdv}$,
are small but not negligible. The radial transport of energy ($Q_{\rm
j}$) occurs in the region of strong gradient of $T_c$. Advection of
energy and adiabatic compression of the gas occur mainly in a region
corresponding to the outflow of gas, i.e. where the gas which carries
a substantial amount of heat ($Q_{\rm adv}$) is compressed as it
enters colder regions of the disc ($Q_{\rm pdv}$).

\subsection{Cooling front velocity}

\begin{figure}
\epsfysize=8.5cm
\epsfxsize=8.5cm
\begin{displaymath}
\epsfbox{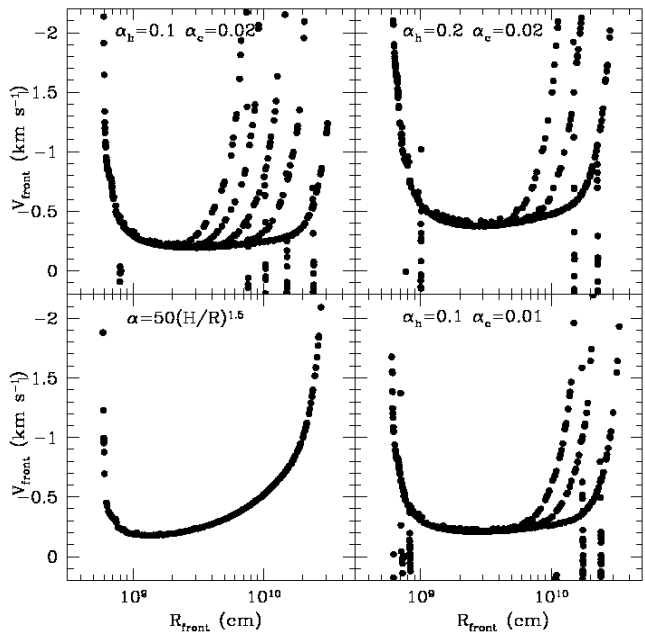}
\end{displaymath}
\caption{The four panels show the speed of successive cooling fronts
in our four models with various $\alpha$-prescriptions (dots). Cooling
fronts appearing at various radii in the outer disc quickly converge
to the same asymptotic speed (in a given model) as they propagate
inward. (Note that all cooling fronts appear at the same location in
model {\it 50hr1.5}.) The asymptotic speed of a cooling front slightly
depends on the value of $\alpha_{\rm hot}.$}
\label{fig:coolspeedmult}
\end{figure}


Figure~\ref{fig:coolspeedmult} shows the speed of cooling fronts in
the four models with various $\alpha$-prescriptions. Successive
cooling fronts in an outburst cycle appear at various locations in the
disc in models using an $\alpha_{\rm
hot}-\alpha_{\rm cold}$ prescription. All 
the cooling fronts appear at the same location, close to the
disc's outer edge, in model {\it 50hr1.5}. The initial speed of all
cooling fronts is relatively high but soon, as they propagate
inward, they relax to a reduced and common asymptotic speed which
depends only on the current location of the front in the disc, $R_{\rm
front}$, in agreement with early calculations by Cannizzo (1994).

The effect of varying $\alpha_{\rm cold}$ on the (asymptotic) speed of
a cooling front is small (at most). Increasing the value of $\alpha_{\rm
hot}$ results, however, in higher asymptotic speeds.

In the asymptotic regime, $V_{\rm front}$ has a weak dependence on
$R_{\rm front}$. This is especially true in models with an
$\alpha_{\rm hot}-\alpha_{\rm cold}$ prescription (see also Cannizzo
1994). The different variation of $V_{\rm front}$ with $R_{\rm front}$
in model {\it 50hr1.5} can be accounted for by the variation of
$\alpha$ with radius in this specific model, according to the scaling
of $V_{\rm front}$ versus $\alpha_{\rm hot}$ in the three other
models. Note that the sharp increase of $V_{\rm front}$ close to the
disc inner edge observed in all models is a boundary condition effect:
the surface density $\Sigma$ goes to zero at $R_{\rm in}$ and this
causes the front to propagate faster as $\Sigma$ is closer to
$\Sigma_{\rm min}$ in the innermost regions of the disc.

The speed of a cooling front, at a given radius $R_{\rm front}$, is
slightly sensitive to the mass of the central object. We obtained an
increase of $V_{\rm front} \sim 30 \%$ in the asymptotic regime when
the mass $M_1$ was changed from $1.2~M_{\odot}$ to $0.6~M_{\odot}$.

$V_{\rm front}$ in the asymptotic regime is not much
larger than the speed of the hot gas into which the front penetrates; 
both are of order of a fraction of $1~{\rm km~s^{-1}}$. This is
consistent with the idea that an evolution of the inner hot disc on a
viscous timescale is typically required to allow the propagation of
the cooling front (Lin et al. 1985, Vishniac \& Wheeler 1996). But
contrary to the case of heating fronts, we find no good correlation
between the value of $V_{\rm r,max}$ inside the transition region and
the speed of cooling fronts.

Recently, Bobinger et al. (1997) directly determined the speed of a
cooling front with the eclipse mapping technique. They observed the
evolution of the brightness temperature profile in the disc of the
dwarf nova IP Peg during four successive days of its decay from
outburst. They infer a mean value of $V_{\rm front} \sim 0.8~{\rm
km~s^{-1}}$ over the four days, which is consistent with the speeds
shown in Fig.~\ref{fig:coolspeedmult}.  It will be difficult, however,
to determine the viscosity parameter $\alpha$ from such observations,
even if variations of $V_{\rm front}$ with $R_{\rm front}$ were
detected. For instance, models {\it h0.1.0c0.02} and {\it 50hr1.5}
produce very different outburst cycles which could be distinguished
from observed lightcurves; yet the variation of $V_{\rm front}$ with
$R_{\rm front}$ is not drastically different. A strong prediction of
the DIM (which probably holds for any reasonable
$\alpha$-prescription) is, however, the rapid decrease of $V_{\rm
front}$ with $R_{\rm front}$ shortly after the appearance of a cooling
front. The detection (or non-detection) of this characteristic would
be strong evidence (or disproof) that the DIM -- as we understand it
-- operates in accretion discs.

Vishniac \& Wheeler (1996) proposed an analytical model of cooling
fronts which uses the numerical results of Cannizzo et al. (1995) as
calibration. In the derivation of Vishniac \& Wheeler (1996), the speed
of a cooling front does not depend on the physical conditions in the
cold gas behind the front. This is in agreement with the independence
of $V_{\rm front}$ with $\alpha_{\rm cold}$ observed in our
simulations.

Vishniac \& Wheeler (1996) predict a speed for the cooling front
\begin{equation}
V_{\rm front} = \alpha_F c_F \left( \frac{H}{R} \right)^{0.54}
\label{eq:vfrontvw1}
\end{equation}
if $\alpha=50 (H/R)^{1.5}$ (and assuming a Kramer's law opacity is
valid for the hot gas), and a speed
\begin{equation}
V_{\rm front} = \alpha_{\rm hot} c_F \left( \frac{H}{R} \right)^{7/10}
\label{eq:vfrontvw2}
\end{equation}
in the case of an $\alpha_{\rm hot}-\alpha_{\rm cold}$ prescription.  Here
$\alpha_F$ and $c_F$ are the viscosity parameter and the sound speed
at the cooling front (i.e. where rapid cooling sets in).

Equation~(\ref{eq:vfrontvw1}) predicts a $V_{\rm front}$ which is
slightly smaller ($20 \%$ typically) than what we find in the
asymptotic regime of model {\it 50hr1.5} but the dependence with
$R_{\rm front}$ appears good. Equation~(\ref{eq:vfrontvw2}) predicts,
however, a $V_{\rm front}$ which is typically twice as small as what
is observed in the asymptotic regime of model {\it h0.1.c0.02}.

Vishniac \& Wheeler (1996) assume that the gas velocity at the cooling
front, $V_F$, is $\sim 1/6~\alpha_F c_F$ and that $V_F$ is much larger
than the cooling front velocity.  In our simulations, $V_F \sim
1/7~\alpha_F c_F$, but the cooling front velocity is not much smaller
than $V_F$. The ratio of the gas velocity at the cooling front to the
cooling front velocity is typically of order 2 (see
Figs.~\ref{fig:cool} and \ref{fig:coolspeedmult}). Neglecting $V_{\rm
front}$ is probably one of the reason for the discrepancy between the
predictions of Vishniac \& Wheeler (1996) and our results.

\subsection{Width of cooling fronts}

Figure~\ref{fig:dwmult1.2} shows the width of cooling fronts in the
four models with various $\alpha$-prescriptions.  Cooling fronts are
much broader than heating fronts but their width $\delta w$ remains
proportional to $H$.  As for heating fronts, we interpret the width of
a cooling front as the distance traveled by the front during a few
thermal timescales (the thermal structure of the front being dominated
by radiative cooling): $\delta w$ is proportional to $R_{\rm
front}^{3/2} \propto H$ (Eq.~(\ref{eq:deltaw})).
\footnote{Naively, one would expect from Eq.~(\ref{eq:deltaw}) smaller
$\delta w$ for cooling fronts than for heating fronts (because of
smaller $V_{\rm front}$). Equation~(\ref{eq:deltaw}) does not take
into account, however, several complications in the problem. The
thermal timescale $\tau_{\rm th} \equiv (\alpha \Omega_K)^{-1}$, which
is very different for annuli in the hot and the cold regions of the
disc, is probably increased because of the ionization energy in
transition regions (Papaloizou \& Pringle 1985). The precursor region
drives out of thermal equilibrium the disc annuli entering the
transition region in the case of heating fronts (see
Fig.~\ref{fig:scurve}), while there is no equivalent for cooling
fronts. Finally, radiative cooling clearly does not dominate the
thermal structure of a cooling front as much as viscous heating does
in a heating front, which probably leads to comparatively longer
thermal timescales for cooling fronts.  A key feature that
Eq.~(\ref{eq:deltaw}) includes, however, is the strong variation of
the thermal timescale with radius.}

This conclusion is in contradiction with the results of Cannizzo et
al. (1995) who find $\delta w = \sqrt{H R}$ in a numerical model using
$\alpha=50 (H/R)^{1.5}$. The profiles of $\delta w/R_{\rm front}$ as a
function of $R_{\rm front}$ shown in Fig.~\ref{fig:dwmult1.2} for
model {\it 50hr1.5} are consistent with those of Cannizzo et
al. (1995). Cannizzo et al. (1995) claim that $\delta w = \sqrt{H R}$
is also true in models using an $\alpha_{\rm hot}-\alpha_{\rm cold}$
prescription. Our three models with $\alpha_{\rm hot}-\alpha_{\rm
cold}$ prescriptions rule out this possibility and show that $\delta w
\propto H$. As mentioned previously, we interpret the flatter profiles
of $\delta w$ in model {\it 50hr1.5} as due to the variation of
$\alpha$ with radius in this particular model. Models using $\alpha=
\alpha_0 (H/R)^{n}$ with various $n$ produce various slopes for the
profiles of $\delta w$ as a function of $R_{\rm front}$.

A comparison of the various panels in Fig.~\ref{fig:dwmult1.2} shows
that the effect of varying $\alpha$ on the width of a cooling front is
small.

\subsection{Self-similar solutions}

Vishniac (1997) proposed a self-similar solution for the structure of
the hot inner disc during the propagation of a cooling front. We
confirm the existence of a self-similar regime, although the
self-similarity described here is somewhat different from the solution
of Vishniac (1997) because $\Sigma$ is found to scale naturally with
$\Sigma_{\rm min}$.  We leave a detailed analysis of the
self-similar solution for a future paper, but we give here our main
qualitative results.

\begin{figure}
\epsfysize=10cm
\epsfxsize=8cm
\begin{displaymath}
\epsfbox{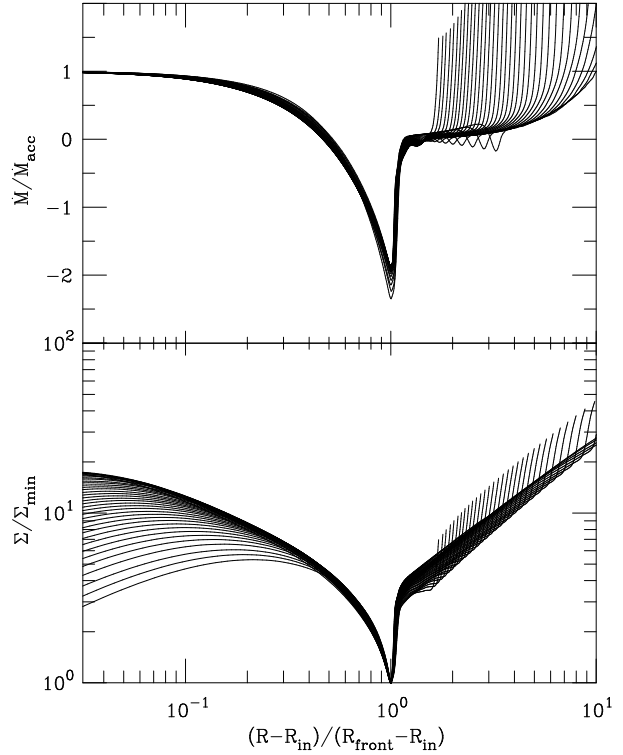}
\end{displaymath}
\caption{Successive normalized profiles of $\Sigma$ and local mass
  flow rate $\dot M$ as a function of the dimensionless radius
  $(R-R_{\rm in})/(R_{\rm front}-R_{\rm in})$ during the propagation
  of a cooling front (model {\it h0.1.c0.02}). The superposition of
  profiles reveals the nearly self-similar evolution of the inner hot
  disc.}
\label{fig:selfsim}
\end{figure}


Figure~\ref{fig:selfsim} shows successive profiles of $\Sigma$ and
local mass transfer rate $\dot{M}$ during the propagation of a cooling
front in model {\it h0.1.c0.02}, when the asymptotic regime is
reached. The profiles of $\Sigma$ have been normalized to the minimum
surface density in the cooling front, which is extremely close to
$\Sigma_{\rm min}$ at $R_{\rm front}$. This is not unexpected, since
the transition from the hot to the cold state occurs precisely when
$\Sigma$ reaches $\Sigma_{\rm min}$. The mass flow rate $\dot{M}$ has
been normalized, following Vishniac \shortcite{v97}, to the mass
transfer rate at the disc inner edge, $\dot{M}_{\rm acc}$.  The
horizontal coordinate chosen here is $(R-R_{\rm in})/(R_{\rm
front}-R_{\rm in})$ in order to minimize the effect of the inner
boundary condition on the profiles. The value of $R_{\rm front}$
varies by more than an order of magnitude for the successive profiles
shown in Fig.~\ref{fig:selfsim} and yet the normalized $\Sigma$ and
$\dot{M}$ remain remarkably similar: except for the effect of boundary
conditions, the evolution of the inner hot disc is close to
self-similar.

The mass accretion rate onto the compact object, $\dot{M}_{\rm acc}$,
is proportional to $\dot{M}_{\rm front}$ at the transition radius. But
\begin{equation}
\dot{M}_{\rm front} = 2 \pi R_{\rm front} \Sigma(R_{\rm front})
V_r(R_{\rm front}),
\end{equation}
$\Sigma(R_{\rm front}) = \Sigma_{\rm min}(R_{\rm front}) \propto
R_{\rm front}^{1.1}$ (e.g. HMDLH), and $V_r(R_{\rm front})$ is
comparable to the front velocity.  If $\alpha$ in the hot inner disc
is constant, one expects a weak dependence of $V_r(R_{\rm front})$
with $R_{\rm front}$ and $\dot{M}_{\rm acc} \propto R_{\rm front}^n$,
with $n$ slightly larger than 2. Indeed, numerical simulations show
that $n \simeq 2.2$ for an $\alpha_{\rm hot}-\alpha_{\rm cold}$
prescription, and $n \simeq 2.4$ when $\alpha \propto (H/R)^{1.5}$.

As discussed by Vishniac \shortcite{v97}, the hot inner disc evolves
on a viscous time during the propagation of a cooling front, and is
therefore very close to thermal equilibrium. The inner disc empties as
a result of accretion onto the compact object, but also because an
even larger amount of mass is transferred into the cold outer regions
of the disc. These parts of the disc are essentially frozen on the
time scale of propagation of the cooling front. Consequently,
$\Sigma(R)$ behind the front remains constant in time and is equal to
$K \Sigma_{\rm min}(R)$, where $K$ is a self-similarity constant that
depends on the mass of the primary and the viscosity parameter
$\alpha$ both in the hot and the cold regions of the disc. For the
parameters of Fig.~\ref{fig:selfsim}, $K \sim 4$. The constant $K$
increases with the mass of the primary (for example, $K \sim 6-7$ for
a 7 M$_\odot$ primary typical of a black hole SXT).

The sudden increase of $\Sigma$ just after the passage of a cooling
front is actually smaller in the pre-self-similar regime, so that
$\Sigma / \Sigma_{\rm min}$ increases as the front propagates toward
the asymptotic regime. This accounts for the density profile in the
outer disc which is relatively flat close to the outer edge, steepens
at smaller radii and finally becomes proportional to $\Sigma_{\rm min}
\propto r^{1.1}$ in the self-similar regime.

\subsection{Reflection of a cooling front}

The sudden increase of $\Sigma$ just after the passage of a cooling
front cannot be arbitrarily large (Vishniac, 1997). If $\Sigma >
\Sigma_{\rm max}$ behind the cooling front, an inside-out heating
front develops and propagates over some distance in the disc because
$\Sigma$ is substantially larger than $\Sigma_{\rm min}$ in the outer
disc. The initial cooling front soon disappears as hot gas from the
outer regions of the disc accretes inward.

This situation corresponds to the reflection of a cooling front into a
heating front and it occurs whenever $\Sigma_{\rm max}/\Sigma_{\rm
min}$ is too small. The ratio $\Sigma_{\rm max}/\Sigma_{\rm min}$ has
a weak radial dependence ($R^{0.02 - 0.03}$; e.g. HMDLH). If the
self-similarity constant $K$ is too large, the cooling front will
experience a reflection typically when it enters the self-similar
regime: the gradual increase of $\Sigma / \Sigma_{\rm min}$ preceding
the self-similar regime will drive $\Sigma$ behind the front above
$\Sigma_{\rm max}$ and lead to the reflection.

The multiple reflections of cooling fronts are well known if
$\alpha_{\rm cold} = \alpha_{\rm hot}$ \cite{sma84}. In this case, the
inner disc can never be brought entirely into quiescence, and the
reflected transition fronts propagate back and forth in a restricted
region of the disc, causing small amplitude quasi-oscillations of the
disc luminosity.

Note that because $K$ increases with the mass of the primary, the
reflections of cooling fronts are more likely to occur in BH SXTs than
in neutron star SXTs or dwarf novae.

\subsection{Exponential decays}

We reexamine, in light of our results on cooling fronts, the claim by
Cannizzo et al. (1995; see also Vishniac \& Wheeler 1996) that
$\alpha=\alpha_0 (H/R)^{1.5}$ is necessary to account for exponential
decays of soft X-ray transients (SXTs; but see also the similarity
between SXTs and dwarf novae pointed out by Kuulkers, Howell \& van
Paradijs 1996)

Our simulations show that the inner accretion rate $\dot M_{\rm acc}
\propto R_{\rm front}^n$ in the self-similar regime of propagation of
a cooling front ($n = 2 - 2.5$ depending on the
$\alpha$-prescription).

Since X-rays are produced close to the central object in an SXT, the
exponentially decaying lightcurves imply an exponential decay of
$\dot M_{\rm acc}$ as well, i.e.
\begin{equation}
\frac{d \dot M_{\rm acc}}{dt} \propto \dot M_{\rm acc}.
\end{equation}
If we assume that the decay phase corresponds to the propagation of a
cooling front in the disc and that our results apply (no irradiation
effect is included in our calculations), then
\begin{equation}
\frac{d \dot M_{\rm acc}}{dt} \propto R_{\rm front}^{n-1} V_{\rm front} \propto \dot M_{\rm acc} \frac{V_{\rm front}}{R_{\rm front}},
\end{equation}
and an exponential decay is possible only if $V_{\rm front}/R_{\rm front}$
remains constant, meaning that $R_{\rm front}$ must also vary exponentially
with time:
\begin{equation}
V_{\rm front} = \frac{d R_{\rm front}}{dt} \propto R_{\rm front}.
\label{eq:exp}
\end{equation}
Note that this requirement was already deduced,
although differently, by Cannizzo et al. (1995) and Vishniac \&
Wheeler (1996).

Figure~\ref{fig:coolspeedmult} shows that $V_{\rm front}$ is primarily
independent of $R_{\rm front}$ (in the asymptotic regime) in the three
models with $\alpha_{\rm hot}-\alpha_{\rm cold}$ prescriptions. On the
contrary, model {\it 50hr1.5} shows a variation of $V_{\rm front}$
with $R_{\rm front}$ which is linear in first approximation. Our
calculations therefore confirm that a dependence of the viscosity
parameter $\alpha$ with radius (close to $\alpha \propto R^{3/4}$ like
in model {\it 50hr1.5}) is required to obtain exponential decays of
$\dot M_{\rm acc}$ in the DIM.

A major concern however is the relevance of interpreting exponentially
decaying lightcurves as due to the propagation of a cooling front in
the discs of SXTs. In particular, Shahbaz et al. (1998) argue that
both linear and exponential decays are observed in SXTs (see also the
compilation of lightcurves by Chen et al. 1997).  Since there is no
obvious physical reason why $\alpha$ would vary markedly from one
system to another, the possibility that both linear and exponential
decays are observed in SXTs argues against an interpretation of the
exponential decays as due to a specific functional form of $\alpha$.


An alternative and perhaps more promising explanation for the
exponential decays has been proposed by King \& Ritter (1998). The
authors argue that strong irradiation of the disc may prolongate the
phase during which the entire disc is in the hot state and lead to an
evolution of the disc on a (long) viscous time (see however
Dubus et al. 1998).

\subsection{Cooling fronts and the periodicity of outburst cycles}

The asymptotic regime of cooling front propagation has a profound
influence on the outburst cycles experienced by the disc. Since the inner hot
disc evolves in a self-similar way, the profiles of $\Sigma$ and $T_c$ inside
$R_{\rm front}$ are uniquely determined by the current value of $R_{\rm
front}$.

A cooling front following a heating front that reached the outermost
regions of the disc therefore propagates in the exact same way a
previous cooling front starting from $R_{\rm out}$ did. This type of
cooling front erases the history of the disc and is responsible for
the periodicity of the outburst cycles in the simulations. An outburst
cycle ends and a new cycle begins when a heating front reaches $R_{\rm
out}$ (also corresponding to the largest amplitude outbursts).

Observed outburst cycles in dwarf novae are usually regular but
not periodic (e.g. Warner 1995, Cannizzo \& Mattei 1992). This may hint
that a piece of physics is missing in the standard DIM. A variation of
the mass transfer accretion rate $\dot{M}_T$ appears as a natural
candidate for explaining the irregularity of observed cycles.

\section{Evolution around S-curves}

\begin{figure}
\epsfysize=8.5cm
\epsfxsize=8.5cm
\begin{displaymath}
\epsfbox{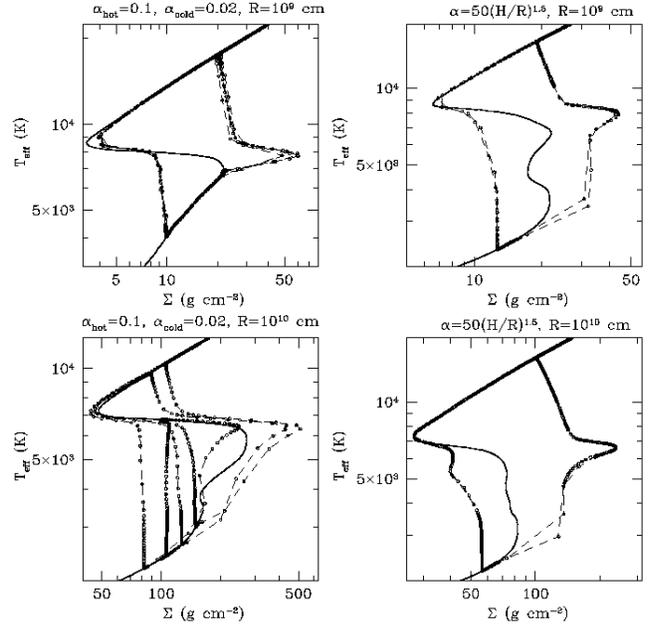}
\end{displaymath}
\caption{Limit cycles experienced by disc annuli located at $R=10^{9}$
and $10^{10}$ cm in the $\Sigma-T_{\rm eff}$ plane. The left panels
correspond to model {\it h0.1.c0.02} and the right ones to model {\it
50hr1.5}. The solid lines are the thermal equilibrium curves, where
$Q^+ = Q^-$ (standard S-curves). The dashed lines trace the successive
states of a disc annulus, while the density of circles is
representative of the amount of time spent by an annulus in a given
state. The evolution of the disc annulus shown in the lower left panel
is complex because this annulus is located in a region of front
reflections.}
\label{fig:scurve}
\end{figure}


Figure~\ref{fig:scurve} shows limit cycles experienced by two disc annuli in
models {\it h0.1.c0.02} and {\it 50hr1.5} in the $\Sigma-T_{\rm eff}$ plane,
where $T_{\rm eff}$ is the effective temperature of the disc.

Global effects (i.e influence by neighboring annuli) appear when an annulus
no longer lies on the S-curve. The passage of a cooling front corresponds to
a jump from the upper hot branch of the S-curve to the lower cold branch.
Inversely, the passage of a heating front corresponds to the jump from the
lower branch to the upper branch. Note that the maximum value of $\Sigma$
reached by a disc annulus on the lower cold branch of the S-curve during a
limit cycle is indicative of the proximity of this annulus to the ignition
radius in the disc.

Figure~\ref{fig:scurve} shows the evolution of disc annuli during the
passage of a transition front: a sudden increase of $\Sigma$ is as
good a signature of the transition as a change of $T_{\rm eff}$ is.
For a cooling front, the sudden change of $\Sigma$ shortly precedes
the phase of rapid cooling of the annulus. For a heating front, the
sudden change of $\Sigma$ corresponds to the passage of the spike of
$\Sigma$. This shows that inside a transition front, the viscous time
is not much larger but comparable to the thermal time since surface
density and effective temperature vary on similar timescales.

The complexity of the limit cycles experienced by the annulus at
$R=10^{10}$ cm in model {\it h0.1.c0.02} is due to the many
reflections of inside-out heating fronts into cooling fronts around
$10^{10}$ cm in this model. Note that the track followed by the
annulus states during the successive jumps from the upper branch to
the lower branch do not cover each other in that case because cooling
fronts have not yet reached the asymptotic regime at $10^{10}$ cm in
model {\it h0.1.c0.02}.

Note that the evolutions around S-curves shown here differ
significantly from the results of Ludwig, Meyer-Hofmeister \& Ritter
(1994) using the theory of infinitely thin transition fronts (Meyer
1984). Some additional limitations of the theory were pointed out by
Ludwig \& Meyer (1998).

\section{Validity of the disc equations for the transition fronts}

It is standard in the theory of thin accretion discs to assume that
the vertical structure of the disc is decoupled from its radial
structure because the former evolves on much shorter timescales than
the latter (e.g. Frank et al. 1992). Because $\Sigma$ and $T_{\rm
eff}$ vary on similar timescales, this assumption is no longer valid
inside transition fronts: the vertical structure is coupled to the
radial structure of the disc. In particular, radial transport of
energy becomes important in the transition regions, while the
transport occurs only vertically in steady thin discs (Frank et
al. 1992).

A correct treatment of the coupling between the radial and vertical
structures of the disc would probably involve 2D simulations. For
simplicity, however, the radial transport of energy is usually
parameterized in a simple manner in 1D disc equations (see also the
approximations of decoupling used for the determination of the cooling
rate $Q^-$, HMDLH). The fact that various prescriptions exist for $J$
(Eq.~(\ref{eq:heat})) shows that the radial transport of energy in the
disc is not well known, and surely not within a factor of a few.

\begin{figure}
\epsfysize=8.5cm
\epsfxsize=8.5cm
\begin{displaymath}
\epsfbox{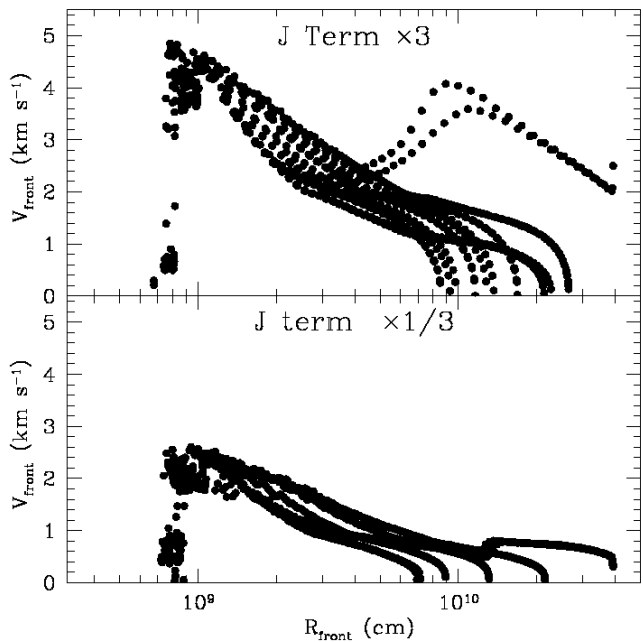}
\end{displaymath}
\caption{An illustration of the sensitivity of the propagation
velocity of heating fronts to the poorly known efficiency of radial
transport of energy (term J in Eq.~(2)), in model {\it h0.1.c0.02}.}
\label{fig:rad3speed}
\end{figure}


Figure~\ref{fig:rad3speed} shows the effect on the speed of heating
fronts of increasing or decreasing by a factor $3$ the intensity of
the term $J$. The large uncertainty in $J$ translates into a
significant uncertainty on the speed of heating fronts. (Note that the
speed of cooling fronts is not affected because the radial transport
of energy has a minor influence on their structure; cf
Fig.~\ref{fig:cool}).

An additional concern about the validity of the disc equations inside
transition fronts comes from the narrowness of the fronts. Lin et
al. (1985) pointed out that having a front width $\delta w \sim H$,
although it does not affect much the Keplerian character of the flow,
could lead to a Rayleigh-unstable situation in the disc. A comparison
between the criterion derived by Lin et al. (1985) and the structure
of the narrow heating fronts found here suggests that this type of fronts
may indeed experience the development of a Rayleigh
instability. This could result in an enhanced radial transport (Lin et
al. 1985) and an additional uncertainty in $J$.

The equations used in this study are only valid in the thin disc
approximation, i.e. when radial pressure gradients can be neglected
and both $\Omega$ and $d\Omega /dR$ are equal to the Keplerian values.
The effects of deviations from Keplerianity and radial pressure
gradients were shown, for instance, by Ludwig \& Meyer (1998) using
full hydrodynamical disc equations and by Kley \& Lin (1996) in
boundary layer computations. Although these effects become
non-negligible in narrow heating fronts, we do not expect our main
results on the structure and properties of transition fronts to be
affected if they were fully taken into account. This could be checked
by taking the full equations, which is beyond the scope of the present
paper.

Finally, we note that the use of a local $\alpha$-prescription in
narrow transition regions ($\delta w \sim H$) is also subject to
caution because the approach of scale separation becomes much less
rigorous on scales so close to the size of the ``turbulent eddies''.

\section{Discussion}

Our simulations show that transition fronts systematically
experience a deceleration during their propagation in the disc,
whatever their direction of propagation. Yet, the front widths
increase with radius in all the cases. This shows that the speed of a
front is not directly related to its width (see also a related claim
by Vishniac \& Wheeler 1996 for cooling fronts). A description
according to which the width of a transition front, which defines the
gradients and the fluxes in the front, also determines its speed
thus appears oversimplified.

Our simulations reveal a qualitative
agreement with one of the predictions of the theory of infinitely thin
transition fronts (Meyer 1984, 1986; see also Papaloizou \& Pringle
1985): we clearly observe that the speed of transition fronts is
related to the proximity of $\Sigma$ in the disc to the critical
values $\Sigma_{\rm min}$ (for cooling fronts) and $\Sigma_{\rm max}$
(for heating fronts).

Although Ludwig et al. (1994) pointed out that this theory may not be
applicable to relatively broad cooling fronts, it could provide an
explanation for the rapid deceleration of cooling fronts: the values
of $\Sigma$ around the region where the cooling front appears are
close to $\Sigma_{\rm min}$, so that the front speed is initially
high; soon, however, the front enters regions with larger and larger
values of $\Sigma$ as compared to $\Sigma_{\rm min}$
(quasi-Shakura-Sunyaev profile), which results in a gradual decrease
of the speed; later, the front must wait for the inner hot disc to
evolve on a viscous timescale, typically, before $\Sigma$ approaches
$\Sigma_{\rm min}$ in the disc, which explains the small speeds in the
asymptotic regime. On the contrary, the speed of a heating front is
able to remain large during the propagation, possibly because the
profile of $\Sigma$ encountered in the disc is typically in between
$\Sigma_{\rm min}$ and $\Sigma_{\rm max}$ (Cannizzo 1993b), i.e. not
too far from $\Sigma_{\rm max}$ everywhere in the disc. If this is
true, the deceleration of heating fronts is due to the (necessary)
proximity of $\Sigma$ to $\Sigma_{\rm max}$ close to the ignition
radius but not elsewhere in the disc.

We have not considered the structure and properties of transition
fronts close to $R_{\rm out}$ in this paper. Indeed, additional
simulations show that in a small region close to the disc outer edge,
transition fronts are affected in a complex way if $R_{\rm out}$ is
allowed to vary with time. The behavior of the fronts in this region
of the disc may also depend sensitively on the precise choice of the
outer boundary condition and how mass is deposited in the disc.

\section{Conclusion}

In this paper, we have investigated the structure and properties of
transition fronts in thin accretion discs with detailed numerical
calculations.

We showed that heating fronts are very narrow and have complex
structures. They propagate at a speed which depends on the profile of
surface density in the disc, the radial transport term in the energy
equation and the value of the viscosity parameter $\alpha$, but
is typically of order a few ${\rm km~s^{-1}}$.

Cooling fronts are broader, have a simpler structure and have smaller
speeds (of order a fraction of a km s$^{-1}$) than heating fronts. We
found that their width is not equal to $\sqrt{HR}$, but is rather
$\propto H$, the local disc scale height, as for heating fronts.

We confirmed that the structure of the inner hot disc is well
described by a self-similar solution during the propagation of a
cooling front. We proposed such a solution in which the surface
density does not scale arbitrarily but with the value of the critical
surface density $\Sigma_{\rm min}$. The self-similarity of the disc
appears responsible for the periodicity of the outburst cycles in the
simulations.

Since all our models show a deceleration of transition fronts during
their propagation in the disc, the observation of such a deceleration
would constitute a nice confirmation that the thermal-viscous disc
instability, as we understand it, is responsible for the large
amplitude variability of discs around white dwarfs, neutron stars and
black holes.

\section*{Acknowledgments}

We are grateful to Charles Gammie, Jean-Pierre Lasota, Ramesh Narayan,
John Raymond and Ethan Vishniac for useful discussions. We also thank
Guillaume Dubus and Eliot Quataert for comments on the manuscript.
This work was supported in part by NASA grant NAG 5-2837. KM was
supported by a SAO Predoctoral Fellowship and a French Higher
Education Ministry grant. RS is supported by a PPARC Rolling Grant for
theoretical astrophysics to the Astronomy Group at Leicester.

\end{document}